\documentclass[final,3p,times,twocolumn]{elsarticle}
\usepackage{graphicx}
\usepackage{amssymb}

\biboptions{square,comma,numbers,sort&compress}

\newcommand{\grad}{\ensuremath{^\circ}}
\newcommand{\nrex}{N-REX$^{+}$}
\newcommand{\parA}{\textsf{parab.~\#1}}
\newcommand{\parB}{\textsf{parab.~\#2}}
\newcommand{\parC}{\textsf{parab.~\#3}}
\newcommand{\parM}{\textsf{parab.~\#2b}}
\newcommand{\parMC}{\textsf{parab.~\#3b}}
\newcommand{\ellA}{\textsf{elliptic~\#1}}
\newcommand{\ellB}{\textsf{elliptic~\#2}}
\newcommand{\strai}{\textsf{straight}}
\newcommand{\mcs}{\emph{McStas}}

\journal{Nuclear Instruments and Methods in Physics Research
Section A}

\begin{document}

\begin{frontmatter}

\title{Parabolic versus elliptic focusing - Optimization of the focusing design of a cold triple-axis neutron spectrometer by Monte-Carlo simulations}

\author[label1,label2]{A. C. Komarek}
\author[label2]{P. B\"{o}ni}
\author[label1]{M. Braden}
\address[label1]{II. Physikalisches Institut, Universit\"{a}t zu
K\"{o}ln, Z\"{u}lpicher Str. 77, D-50937 K\"{o}ln, Germany}
\address[label2]{Physik Department E21, Technische Universit\"{a}t M\"{u}nchen,
James-Franck Str., D-85748 Garching, Germany}

\begin{abstract}
We present Monte-Carlo simulations for the focusing design of a
novel cold-neutron triple-axis spectrometer to be installed at the
end position of the cold guide NL-1 of the research reactor FRM-II
in Munich, Germany.  Our simulations are of general relevance for
the design of triple-axis spectrometers at end positions of
neutron guides. Using the \mcs\ program code we
performed ray trajectories to compare parabolic and elliptic focusing concepts.
In addition the design of the
monochromator was optimized concerning crystal size and mosaic
spread. The parabolic focusing concept is superior to the elliptic
alternative in view of the neutron intensity distribution as a
function of energy and divergence. In particular, the elliptical
configuration leads to an inhomogeneous divergence distribution.
\end{abstract}


\end{frontmatter}


\section{Introduction}

\label{intro} Neutron focusing techniques have become rather
important in triple-axis spectroscopy because there is a high
demand for measurements which require an intense, focused neutron
beam on small samples. Double focusing at the monochromator side
combined with a horizontally focusing analyzer were introduced by
B\"uhrer \cite{buehrer} and Pintschovius \cite{virts1} in the
1980s and became very successful. The double focusing
monochromator allowed for inelastic neutron studies of the phonon
and magnon excitations with the first available small crystals of
high-temperature cuprate superconductors \cite{cuprates} as well
as for the routine measurement of the phonon dispersion under
pressure up to 10~GPa \cite{goncha,klotzA,klotzB,klotzC}.
Especially the ongoing interest in experiments using high-pressure
cells with very small samples required further experimental
progress which could be achieved by the use of focussing neutron
optics first introduced in front of the sample  by Goncharenkov
\emph{et al.} \cite{goncha-focus}. Using elliptical neutron optics
it became possible to study samples as small as 0.1~mm$^3$ by
neutron scattering techniques \cite{press1,press2,press3}, which
is of particular importance in the case of novel materials like
the new, exciting class of high-temperature superconducting
iron-pnictide compounds \cite{FeAs1,FeAs2,FeAs3,FeAs4}, as it it
is frequently difficult to obtain large samples. Focused neutron
beams are also of general importance for inelastic neutron
scattering experiments since the gain of intensity results in a
decrease of measuring time, in an increase of the
measurement-statistics, and often also in a better signal to noise
ratio.

\par Due to the ongoing progress in neutron optical developments
it is now possible to use supermirror guides with different shape
or tapering designs and rather large angles of reflection of the
supermirror coatings
\cite{smirror3,smirror4,smirror5,smirror1,smirror2}. Together with
the virtual source concept \cite{virts1,virts2} and large focusing
monochromator arrays \cite{mono1,mono2,mono3,mono4,mono5} the
intensity at the sample position can be enhanced significantly.

Recent simulations \cite{schanzer} indicate a flux
gain of the order of five which could be obtained by using an
elliptic focussing guide instead of a conventional guide with an
$m$-value of two ($m$ is a measure of the reflection angle
relative to the angle of reflection of Ni).

\par
Here, we compare elliptic, parabolic and conventional (straight) guide
concepts for a cold-neutron triple-axis spectrometer using the free \mcs\ code \cite{mcstas1,mcstas2,mcstas3}.
Our results show that for an instrument at the
end of a cold neutron guide, the parabolic concept is clearly
superior to the elliptic concept concerning the
intensity-divergence distribution.

\section{The optimized focusing concept for a cold-neutron triple-axis spectrometer}

Our calculations were made to design the concept of a new
cold-neutron triple-axis spectrometer which will be installed at
the end of a curved neutron guide (NL1) at the research reactor
FRM-II in Garching. Therefore we have used the specific dimensions
and coatings as well as the intensity distribution of this neutron
guide, but our main results are of broad relevance for any
cold-neutron triple-axis spectrometer placed at a neutron-guide
end-position. A gap of 400~mm is required for the monochromator of
the upstream instrument. The width and the height of NL-1 (coating
$m$~=~2) amount to 60~mm and 120~mm, respectively. In
Ref.~\cite{ellipt} a fully elliptic guide concept has been
presented and a distinctly larger entrance width (70~mm) of the
guide after the position of the upstream instrument has been
proposed in order to overcome the neutron loss caused by the gap
where the monochromator is hosted.

\begin{figure}[!t]
\begin{center}
\includegraphics*[width=1\columnwidth,clip]{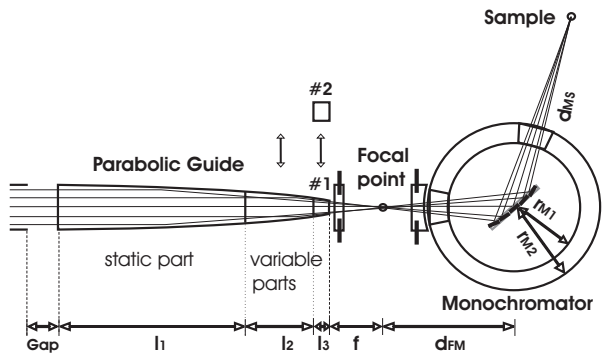}
\end{center}
\caption{Configuration of the cold-neutron spectrometer.
First, a parabolic guide
is installed permanently. The following two parabolic guides
can be removed and replaced by other guide elements.
}
\label{figC}
\end{figure}

\par Fig.~\ref{figC} shows the overall design of the calculated parabolic focussing
concept of the triple-axis spectrometer. The parabolic guide has a
total length $l=l_1+l_2+l_3$ of 6.7~m and is divided into three
sections $l_1$, $l_2$ and $l_3$. The first section $l_1$ has a
length of 5.15~m and is installed permanently. The other two
sections $l_2$ and $l_3$ with 1.4~m and 0.15~m length can be moved
transversely and replaced by other (straight) guide elements of
identical length. The largest section $l_1$ also hosts the
polarizing cavities. Its design will be subject of
a forthcoming publication \cite{VVV}. Several focal distances have
been studied in great detail ranging from 0.1~m to 0.6~m. Finally,
a focal length $f$~=~0.3~m has been chosen as an optimal value
for highest resolution and intensity. After the parabolic guide a
velocity selector will be installed in order to suppress the
higher harmonics and to reduce the background of the instrument.
Additionally, several diaphragms and flight tubes shall be
installed between the parabolic guide and the monochromator drum
for similar purposes (e.g. background reduction). Another
diaphragm at the focal point may also be used as a virtual source
if even higher energy resolution is desired. In the monochromator
drum a double focusing monochromator is installed at a distance
$d_{FM}$ away from the focal point; compare Fig.~\ref{figC}. The
size of the highly oriented pyrolythic graphite (HOPG) crystals
has been chosen as in Ref.~\cite{ellipt}, i.e. with 20~mm $\times$
20~mm size, and the mosaic spread is assumed to be 0.5\grad. All
configurations have been compared based on these identical
conditions. Finally, we will also discuss a different
monochromator design which is optimized for the short distances
$d_{FM}$ and $d_{MS}$~=~1.2~m with $d_{MS}$ being the
distance between monochromator and sample; compare
Fig.~\ref{figC}. Both, the intensity at the sample position and the
peak profile as a function of horizontal divergence and wavelength
could be altered and, hence, designed to our needs by variation of
the distance $d_{FM}$~=~$d_{MS}-\Delta d$. At the sample position
a neutron counter of 20~mm $\times$ 20~mm size has been used in
order to measure the neutron flux at the sample position.

\section{Comparison of different focusing concepts}
Basically, there are six different focusing configurations which will be discussed in this work:\\
\begin{tabular}{ll}
$\bullet$ \parA & parabolic guide ($l_1$ to $l_3$ parabolic) \\
$\bullet$ \parB & parabolic guide ($l_3$ is straight guide) \\
$\bullet$ \parC & parabolic guide ($l_2$, $l_3$ are straight) \\
$\bullet$ \ellA & elliptic guide (all noses; $f$~=~0.3~m) \\
$\bullet$ \ellB & elliptic guide (no noses; $f$~=~1.2~m) \\
$\bullet$ \strai & standard straight guide   \\
\end{tabular} \\
The first configuration, \parA, is a fully parabolic guide with
$f$~=~0.3~m. In the second parabolic configuration, \parB, the
last nose with 0.15~m length is replaced by a standard straight
guide. In \parC\ also the much longer second nose $l_2$ is replaced by a straight guide changing the properties of this concept distinctly.
For the sake of simplicity, the third parabolic concept will be discussed later in section 4.
All parabolic configurations start with 60~mm width of the
entrance window after the gap for the upstream instrument.
\begin{figure}
\begin{center}
\includegraphics*[width=0.9\columnwidth,clip]{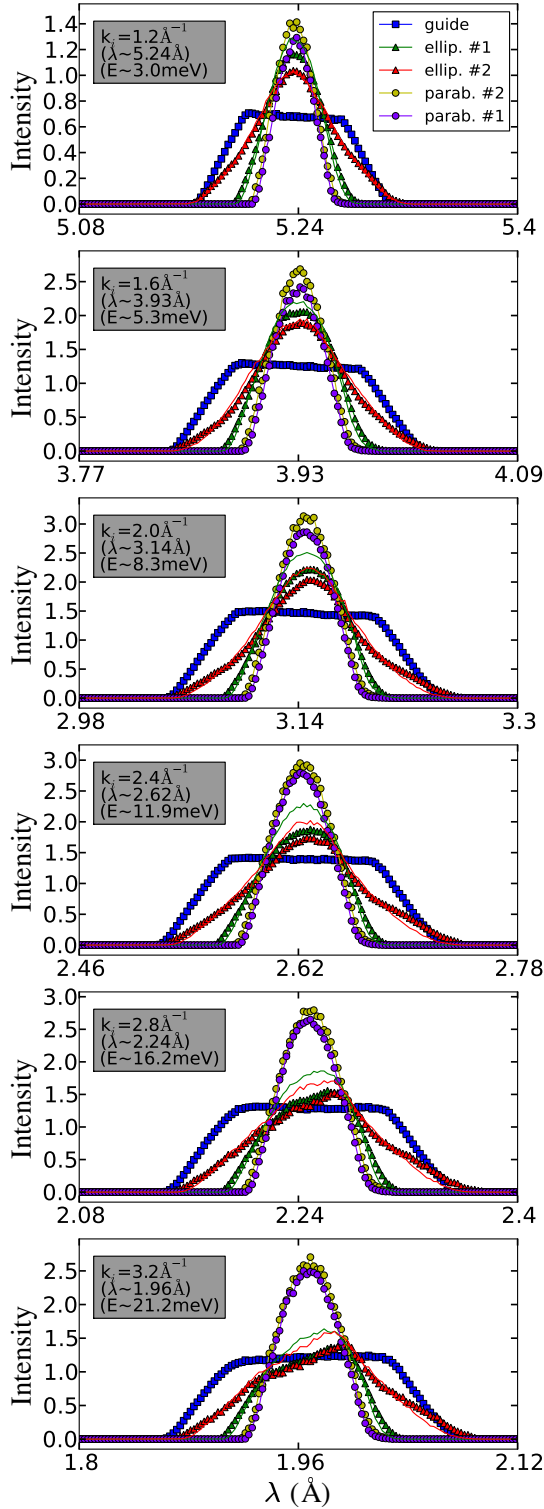}
\end{center}
\caption{Neutron intensity at the sample position as a function of
wavelength for the five different guide concepts (see text).}
\label{figD}
\end{figure}
In contrast, for the two elliptic configurations an increased
entrance width of 70~mm has been applied since this widening of
the guide was recommended in Ref.~\cite{ellipt}. Regarding the two elliptic
setups, all elliptic noses were used in the first
elliptic configuration, \ellA, with $f$~=~0.3~m, and all noses
were removed in the second elliptic configuration, \ellB, yielding $f$~=~1.2~m.
Finally, the configuration \strai\ is a
standard straight guide with length $l$.
The distance between gap and focal point is
identical for all concepts compared in this work.
All guides
have supermirror-coatings with $m$~=~2 and 4 for the top/bottom
and the sides, respectively (see Tab.~\ref{tabA}). However, for
the final parabolic configuration an additional intensity gain of
about 10\% could be achieved by an optimized distribution of the
$m$-value of the supermirror coating along the parabolic guide
with smaller values at the beginning but larger values at the end.
\begin{table}
\centering{ {\footnotesize
\begin{tabular}[t]{lccc}
\hline\hline
  Parameter &  \parA /            &  \ellA /      & \strai    \\
    &   \parB  /          &   \ellB          \\
    &   \parC            &            \\
\hline
Entrance width & 60 mm & 70 mm & 60 mm \\
Exit width & 12.4 mm / & 27.8 mm / & 60 mm \\
           & 15.2 mm /  & 52.8 mm   &     \\
           & 21.6 mm    &     \\
Entrance height & 120 mm & 120 mm & 120 mm\\
Exit height & 120 mm & 120 mm & 120 mm \\
focusing guide length & 6.70 m /    & 6.70 m / &  0 m \\
\emph{(length of parabolic }             & 6.55 m /    & 5.80 m   &     \\
\emph{or elliptic guides) }            & 5.15 m      &     \\
straight guide length & 0 m /    & 0 m / &  6.70 m \\
\emph{(of guides following }             & 0.15 m /    & 0 m   &     \\
\emph{the focusing guides)}             & 1.55 m      &     \\
Coating (hor.) & av. 3.5 & 4 & 4 \\
Coating (vert.) & 2   & 2 & 2 \\
focal distance $f$ \dag            & 0.3 m / & 0.3 m / & $\infty$ \\
                    & 0.45 m / & 1.2 m   & \\
                    & 1.85 m  &     & \\
focal distance $f'$ \ddag            & $\infty$  & 2.5 m  & $\infty$ \\
Distance gap-$F$ $\sharp$ & 7.0~m  & 7.0 m  & 7.0 m \\
Mono. shift $\Delta d$ & -0.3~m & -0.1~m & -0.1~m \\
$d_{MS}$     &    1.2~m & 1.2~m & 1.2~m \\
$R_{HOR}$ & 1.4 / 1.4 / 1.6 & 1.3 & --- \\
 \hline\hline
\end{tabular}
}} \caption{\label{tabA} Guide parameters for the five
different focusing concepts.  \dag: $f$ is the distance from the
end of the last parabolic or elliptic component to the focal
point. Only horizontal focusing has been applied for all models.
\ddag: $f'$ is the focal distance of the entrance of the elliptic
guide. $\sharp$: 'gap-$F$' is the distance between the gap of the
upstream instrument and the focal point $F$. $R_{HOR}$: Horizontal
monochromator curvature prefactor (see text). }
\end{table}
Thus, the guide starts with an $m$-value of 2.5 which continuously
increases up to a value of 3.5 at the end of section $l_1$, i.e.
after 5.15~m length. Only the two short noses $l_2$ and $l_3$
possess coatings with higher $m$-values which continuously
increase from 4 at the beginning of $l_2$ to 6 at the end of nose $l_3$.
Hence, on average, the $m$-value of the
supermirror coating for the parabolic configurations amounts to
3.50. As mentioned before, compared with a uniform coating with
$m$~=~4 the average intensity gain is of the order of 10\%; it is
smaller for lower neutron energies and only slightly higher for
very high neutron energies.
\par Regarding the vertical monochromator curvature the
standard monochromator curvature (as expected for focusing from
focal point to the sample position by the Rowland condition) has
been chosen for all focusing guides. For the horizontal
monochromator curvature a different value has been chosen which
has been optimized by comparison of intensity and resolution for
different horizontal monochromator curvatures in 10\% steps from
100\% of the nominally optimal curvature (Rowland condition) up to
200\% of this nominal value. Finally, a slightly relaxed curvature
around 140\% of the nominal value turned out to be ideal regarding
the intensity. Thus, the standard horizontal curvature has been
multiplied by a factor $R_{HOR}$ which is given in
Tab.~\ref{tabA}.

\begin{figure}[!t]
\begin{center}
\includegraphics*[width=0.95\columnwidth,clip]{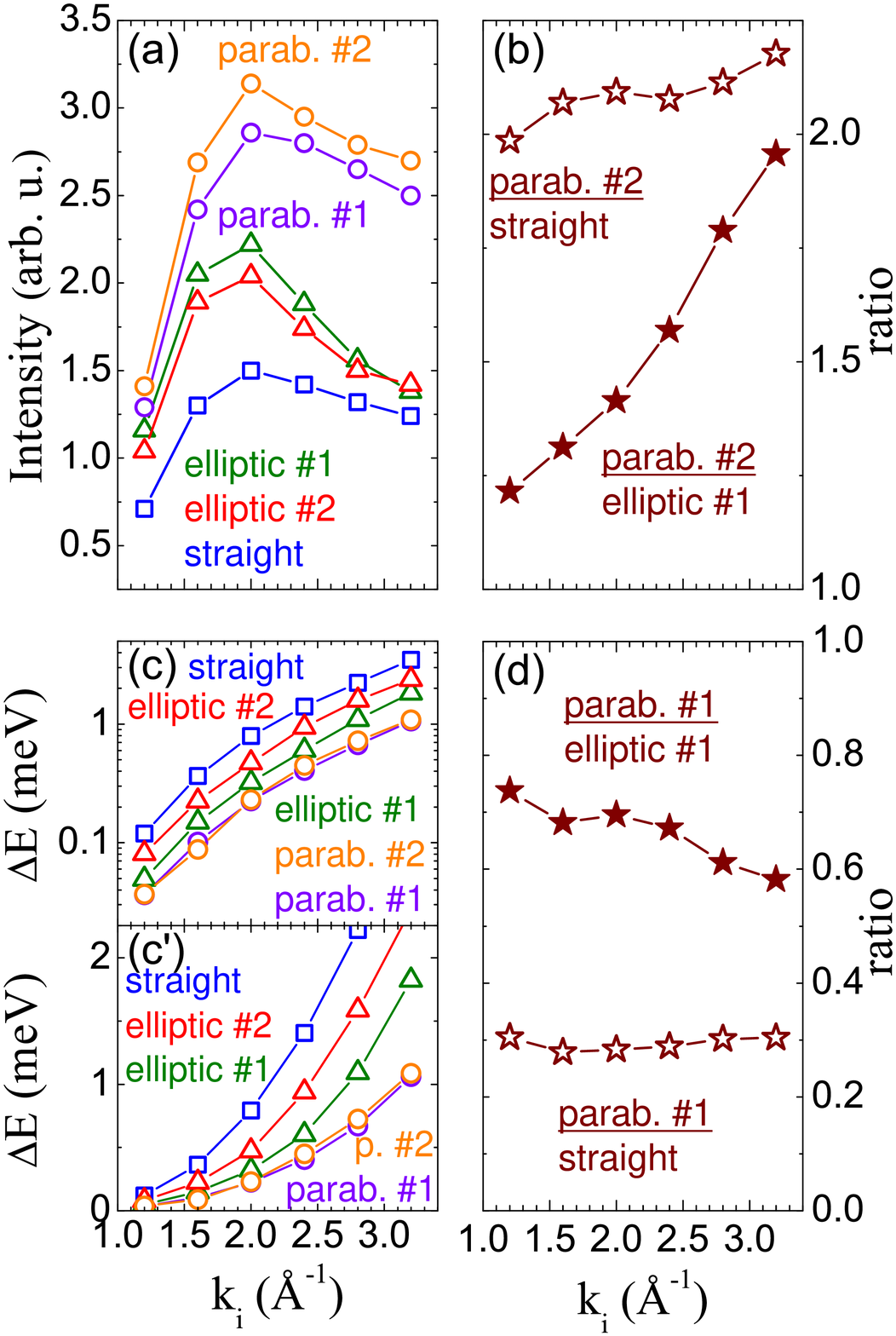}
\end{center}
\caption{(a) Intensity (amplitude) and (c,c') energy resolution
(FWHM) for the five different models
\parA\ (\emph{violet circles}), \parB\ (\emph{yellow-orange circles}),
\ellA\ (\emph{green triangles}), \ellB\ (\emph{red triangles}) and
\strai\ (\emph{blue squares}) as a function of $k_i$. (b) Ratio of
the intensities of \parB\ and \ellA\ (closed
symbols) or \strai\ guide (open symbols). (d) Ratio of the energy
resolution (FWHM) of \parA\ and \ellA\ (closed symbols) or the \strai\ guide (open
symbols).} \label{figB}
\end{figure}
\begin{figure}[!t]
\begin{center}
\includegraphics*[width=0.72\columnwidth,clip]{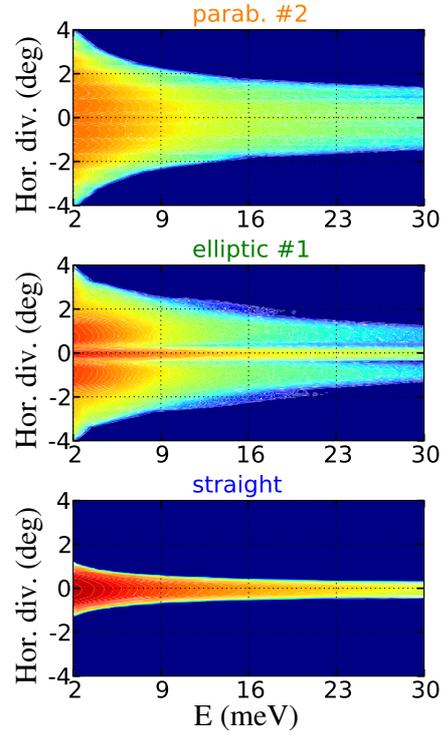}
\end{center}
\caption{Neutron intensity directly after each guide as a function
of energy and horizontal divergence for the \parB, \ellA\ and
\strai\ concept. The intensity $I$ is plotted logarithmically
ranging from $I\leq 10^5$ (\emph{blue}) to $I=10^9$ (\emph{dark
red}).} \label{figN}
\end{figure}
For the elliptic models, the guide parameters from
\cite{ellipt,janosch} have been taken without any further
optimization. (For the other models, all parameters have been
always optimized for highest intensity at the sample position!) In
the elliptic case the total length of the elliptic guide is
defined by the focal distance $f$. For the \ellA\ concept it is
identical to
\parA\ concept and for \ellB\ it is 0.9~m shorter. For the
elliptic cases $f'$ is the other side focal point at the entrance
window; compare also \cite{janosch}.

\subsection{Neutron intensity and energy resolution}
\par Fig.~\ref{figD} shows the  neutron intensity
calculated at the sample position as a function of wavelength for
the five different configurations and for different values of the
incident neutron wavevector $k_i$
ranging from 1.2~\AA$^{-1}$ to 3.2~\AA$^{-1}$.
Additionally, the intensities for \ellA\ and \ellB\ have been calculated but
without the guide widening to 70~mm proposed in
reference \cite{ellipt}, i.e. with the standard entrance window of
60~mm width only (\emph{green} and \emph{red lines} in
Fig.~\ref{figD}). The widened guides are not useful. On the one
hand more neutrons are collected by the wide-spread entrance
window of such a widened guide, but it becomes more difficult to
focus such a beam on a small 2~cm~$\times$~2~cm spot at the sample
position once it is initially widened.
\begin{figure*}[!t]
\begin{center}
\rotatebox{0}{\includegraphics*[width=1.75\columnwidth,clip]{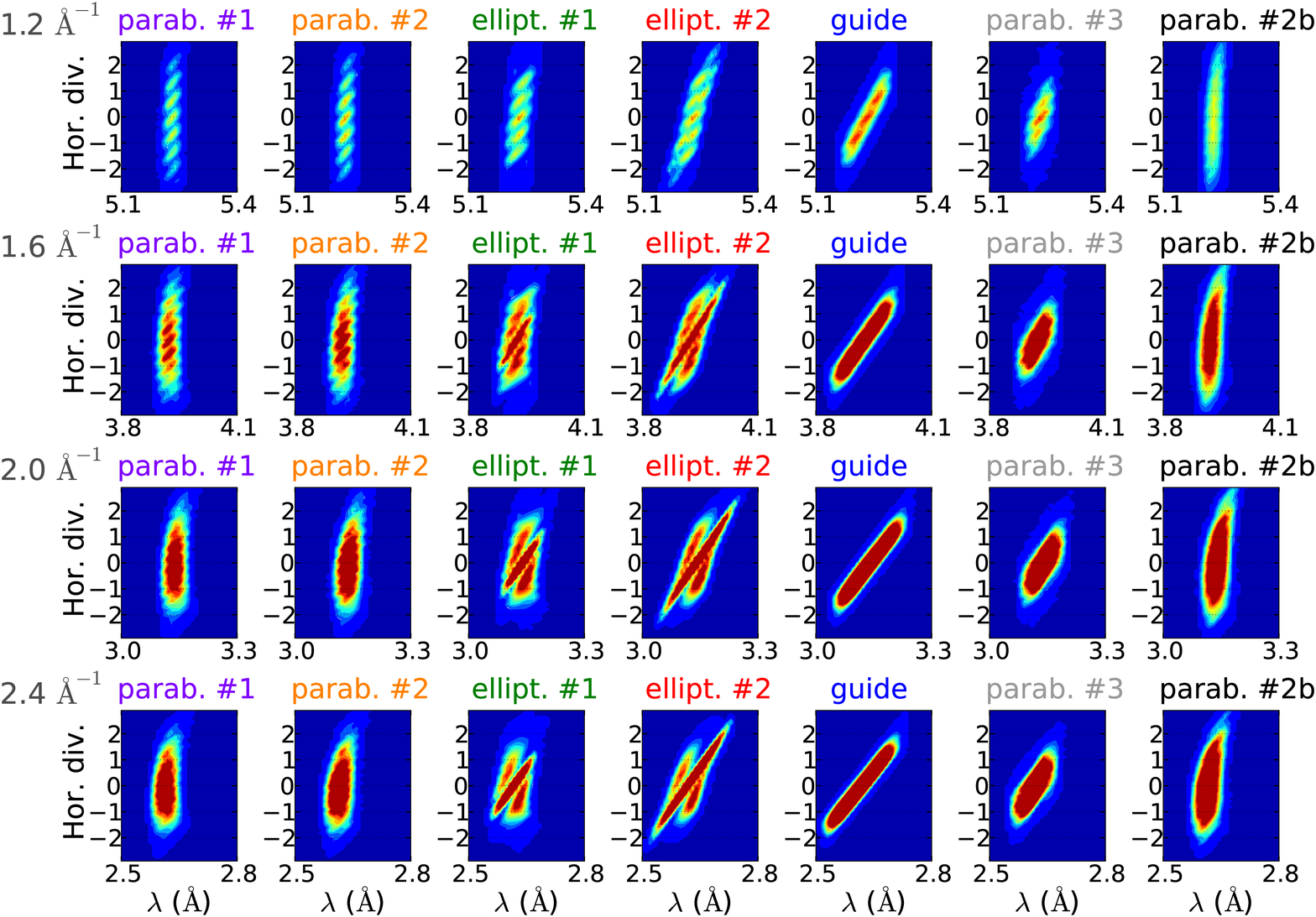}}
\end{center}
\caption{Intensity-ellipsoids at the sample position as a function
of $\lambda$ and hor. div. for the different concepts (see text)
at different values of $k_i$. The intensity is plotted from 0.0
(\emph{blue}) to $3\cdot 10^{6}$ (\emph{dark red}) in arbitrary
units. (\parC\ and \parM\ will be discussed in the sections 3 and 6.)}
\label{figH}
\end{figure*}
\par As can be seen from Fig.~\ref{figD}, the energy resolution of
the two parabolic configurations is distinctly better than that of
the elliptic configurations. Furthermore, the neutron intensity of
the guides with parabolic tapering is distinctly larger, especially
for larger values of $k_i$. In Fig.~\ref{figB} a quantitative
evaluation and comparison of parabolic and elliptic concepts is
presented showing the intensity and energy resolution ratios of
both concepts (\emph{closed stars}). The parabolic concept
performs already better for small incident neutron energies; the
energy resolution, i.e. the peak width (FWHM), of the competing
high-resolution elliptic setup \ellA\ is about one third larger
(worse) than that of the parabolic setup \parA. For higher neutron
energies, the difference is even more dramatic since the energy
resolution, i.e. the peak width (FWHM), of
\ellA\ has become even two thirds larger (worse) than that of the
parabolic setup at a value of $k_i$ equal to 3.2~\AA$^{-1}$
($\sim$21~meV). Furthermore, the neutron intensity at this value
of $k_i$ has become almost two times larger than that of the
elliptic alternative; see Fig.~\ref{figB}.
\par In Fig.~\ref{figB}~(b,d) a comparison of the parabolic concept
with the simple straight guide is shown (\emph{open stars}).
Whereas the parabolic concept is always distinctly better than a
straight guide, the elliptic alternative \cite{ellipt} loses most
of its advantages at higher neutron energies; compare also
Fig.~\ref{figD}.

\begin{figure*}[!t]
\begin{center}
\includegraphics*[width=1.5\columnwidth,clip]{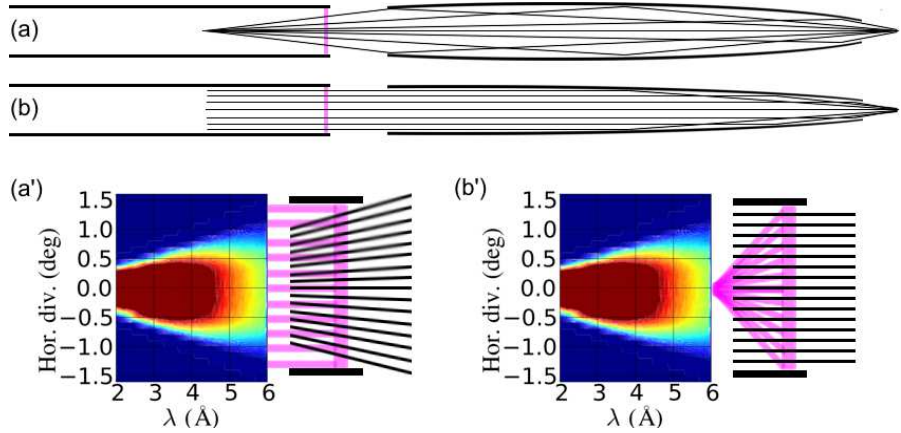}
\end{center}
\caption{Schematic focusing design of (a) fully elliptic and (b)
fully parabolic guides in the horizontal plane. (a',b') The
intensity distributions of neutrons at the entrance window
 are shown as a function of divergence and wavelength together with
a schematic drawing of the entrance window to the right
(\emph{vertical magenta bars}) and the neutron trajectories
(\emph{black lines}) which belong to neutrons that will be focused
into the focal point of the corresponding focusing guide. The
\emph{magenta lines} connecting the schematic drawing of the
entrance window with the contour maps indicate from which
divergence regime in the contour maps these neutrons originate
depending on their position at the entrance window.} \label{figF}
\end{figure*}
\subsection{The transverse Q-resolution}
\par In Fig.~\ref{figN} the neutron intensity is plotted as a function
of horizontal divergence and neutron energy. For
the standard straight guide the horizontal divergence of the
neutrons is rather small compared to the fully focusing
concepts. In the other extreme, the parabolic guide
\parB\ exhibits a broad divergence. From these results, the
straight guide with the worst energy resolution has the best transverse Q-resolution and the parabolic guide with
the best energy resolution has the worst
transverse Q-resolution. \ellA\ looks like a superposition of straight and parabolic concepts.  Thus, the
intensity distribution of the elliptic guide is not as homogenous
as for the parabolic or for the straight guide. Hence,
with elliptic focusing one might expect three major
peaks for a transverse scan in a neutron scattering experiment. In
order to elucidate this question, we have also calculated the
corresponding neutron intensities at the sample position after
scattering by the monochromator.
\par In Fig.~\ref{figH} the neutron scattering intensity at the
sample position is shown as a function of neutron wavelength and
horizontal divergence for the different models mentioned above.
The calculated intensity maps are directly related to the
resolution ellipsoid of the spectrometer. As can be seen in
Fig.~\ref{figH}, both elliptic concepts always exhibit a multiple
peak structure. Therefore, also a perfect single crystal may
exhibit multiple peaks in a transverse scan. In contrast, the
multiple peak structure is absent in the parabolic concept for
medium and high neutron energies. Only at rather low neutron
energies a similar multiple peak structure may appear.
In the parabolic arrangement the multiple peak structure, however,
can be suppressed when replacing the focussing noses by straight guides
(\parC), see Fig.~\ref{figH}.

\subsection{The qualitative understanding of parabolic versus elliptic
configurations}
\par The better performance of the parabolic
versus the elliptic focusing concept for an instrument at the end
of a neutron guide can be understood by the following reason: the
parabolic guide focuses neutrons with small divergence
from the whole entrance window into its focal point, whereas the
elliptic guide focuses neutrons with higher divergence from outer
parts of the entrance window into its focal point. This can be
seen in the schematic drawings of Fig.~\ref{figF}~(a,b). At larger
distances from the reactor, the neutron intensity provided by the
guide is dependent on the neutron divergence. As can be seen in
Fig.~\ref{figF}~(a',b') the intensity is strongly reduced for
neutrons with higher divergence due to the reflection losses in
the preceding neutron guide. This reduction of intensity for
neutrons with higher divergence becomes especially significant for
higher neutron energies. Since the elliptic concept also depends
on focusing neutrons out of these high-divergence regimes into its
focal point, it has a clear disadvantage with respect to the
intensity, especially for higher neutron energies. In contrast,
the parabolic concept basically focuses neutrons with low
divergence (from the regime with  highest flux in the
intensity-divergence distribution) into its focal point.

\section{Modification for High-Q resolution}
With a basic modification of the parabolic concept one may enhance
the Q-resolution. As can be seen in Fig.~\ref{figC}, it is
possible to remove both parabolic noses of the calculated design
($l_2$ and $l_3$) and to exchange them by straight guides. We will
denote this alternative setup by \parC. This parabolic concept is
almost equivalent to the \ellB\ concept. Like for the other
concepts we have optimized the horizontal monochromator curvature
for neutron flux at the sample position.
\par In Fig.~\ref{figH} the intensity ellipsoids of the different
focusing concepts \parA, \parB, \ellA, \ellB, \strai\ and \parC\
are shown for the most relevant values of $k_i$. As mentioned
before, the elliptic concepts exhibit a multi-peak structure for
all incident neutron energies. The parabolic concepts do not show
such a multiple peak structure except at low energy. But in
contrast to the elliptic concept, the parabolic concept loses this
multiple peak structure when all noses are removed, i.e. in the
\parC-design as can be seen in Fig.~\ref{figH}.
Regarding the intensity at the sample position, this concept is
comparable to the elliptic concept in the lower or medium energy
regime but still clearly outperforms the elliptic alternative in
the high energy regime. This configuration will also have a better
Q-resolution as discussed in the next section.

\section{Simulation of measurements}
In order to study the different focusing concepts under
measurement conditions, we have also simulated several types of
scans: 1$^{st}$, energy scans for an
incoherent scatterer, 2$^{nd}$, elastic scans
across a Bragg peak and, 3$^{rd}$, scans
for measuring an acoustic phonon dispersion will be presented.
\par The
double-focusing analyzer was composed of 11 rows and 21 columns of
10~mm $\times$ 20~mm large HOPG crystals with 0.4\grad\ mosaic
spread. The sample-analyzer and analyzer-detector distances are
rather small, i.e. 0.8~m each, since the spectrometer is intended
to have a rather compact design. The width of the detector was set
to 2~cm and all scans have been performed in the constant-$k_f$
mode of operation. Furthermore, no collimation was used at any
point corresponding to an all-open configuration.

\subsection{Energy scan for an incoherent scatterer}
\begin{figure}[!b]
\begin{center}
\includegraphics*[width=1\columnwidth,clip]{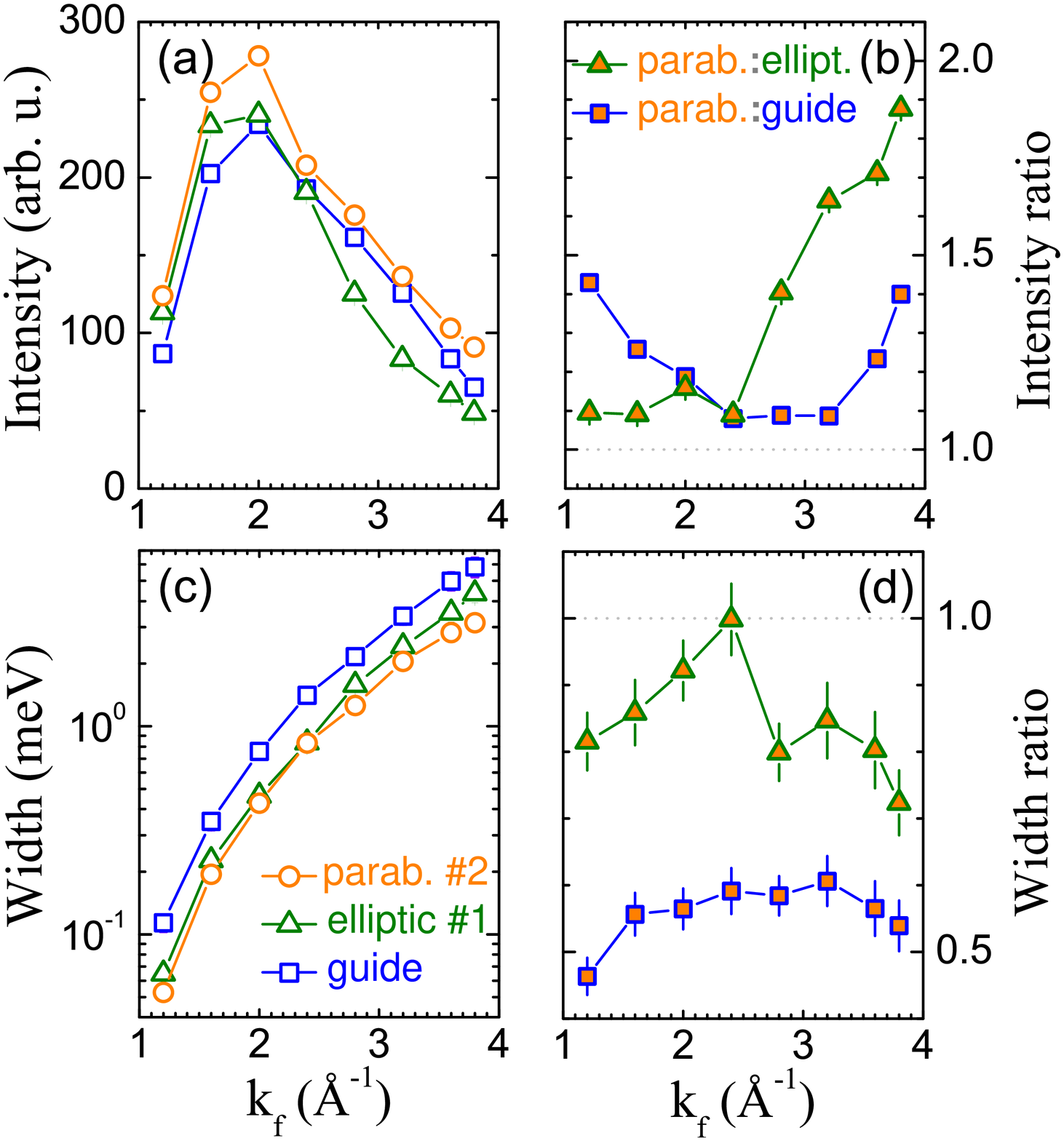}
\end{center}
\caption{(a) Intensity (amplitude) and (c) peak widths (FWHM) of
simulated energy-scans across the elastic line of an incoherent
scatterer; \emph{orange circles}:
\parB, \emph{green triangles}: \ellA, \emph{blue squares}:
\strai. In (b) and (d) the corresponding ratios are shown;
\emph{green/orange triangles}: ratio of \parB\ versus \ellA,
\emph{blue/orange squares}: ratio of \parB\ versus \strai.}
\label{figK}
\end{figure}

First, we have calculated the intensity and
energy resolution for energy-scans across the
elastic line of a simple cylindrical
incoherent scatterer of 20~mm height and 5~mm radius. Each point in reciprocal space was chosen
such that $Q$ was equal to $k_f$ and a double-focusing analyzer
geometry has been chosen. The resulting intensities and peak
widths (FWHM) are shown in Fig.~\ref{figK}.
As can be seen, the intensity of \parB\ is
always larger than the intensity either of the standard straight
guide or of \ellA. At the same time the peak
width of \parB\ is smaller than that for the other
concepts. Compared to the standard straight guide, the energy
resolution is nearly a factor 2 better for \parB.
It is even around 20\% better compared to the high-resolution
elliptic concept \ellA. Especially at very high energy around
30~meV, the resolution of the parabolic concepts becomes almost
30\% better and the intensity 90\% larger when compared to the
elliptic concept.
\par Comparing elliptic and straight guides, the
elliptic guide always has a clearly better energy resolution.
Even the intensity is slightly larger for the elliptic concept at
lower neutron energies. However, the intensity performance changes
at higher neutron energies where the straight guide exhibits a
better performance.

\subsection{Elastic scans}
\begin{figure}[!b]
\begin{center}
\includegraphics*[width=1\columnwidth,clip]{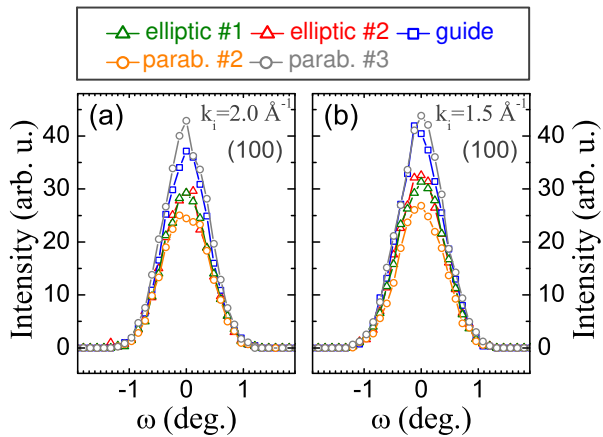}
\end{center}
\caption{Simulation of $\omega$-scans through (1~0~0) at 2~meV
with (a) $k_f$~=~2.0~\AA$^{-1}$ and (b) $k_f$~=~1.5~\AA$^{-1}$
respectively. \emph{Orange circles}: \parB, \emph{green
triangles}: \ellA, \emph{red triangles}: \ellB, \emph{blue
squares}: \strai, \emph{gray circles}: \parC.} \label{figL}
\end{figure}
Next, we have simulated $\omega$-scans across a (100) Bragg reflection of a cubic single-crystal with 5~\AA\
lattice parameter and 50' mosaic spread using a flat analyzer geometry.
But since it is unfavorable to work with a flat
monochromator, if the preceeding parabolic or elliptic guide is
strongly focusing, we did not change the monochromator curvature.
The resulting intensities of such $\omega$-scans are shown
in Fig.~\ref{figL} for two different incident neutron energies.
The highly focusing concept \parB\ exhibits the
lowest performance. But also the highly focusing
alternative \ellA\ is not competitive with a straight
guide. Interestingly, again, the straight guide exhibits
the best intensity-performance for such elastic scans across a
Bragg peak. However by changing towards the
\parC-configuration even higher intensities than for the
straight guide can be obtained; see Fig.~\ref{figL}. Whereas the two very similar
fully and almost fully parabolic configurations
\parA\ and \parB\ are optimized for highest energy resolution,
the 'semi-parabolic' configuration \parC\ is optimized for
Q-resolution.

\subsection{Phonon measurements}
\begin{figure}[!b]
\begin{center}
\includegraphics*[width=1\columnwidth,clip]{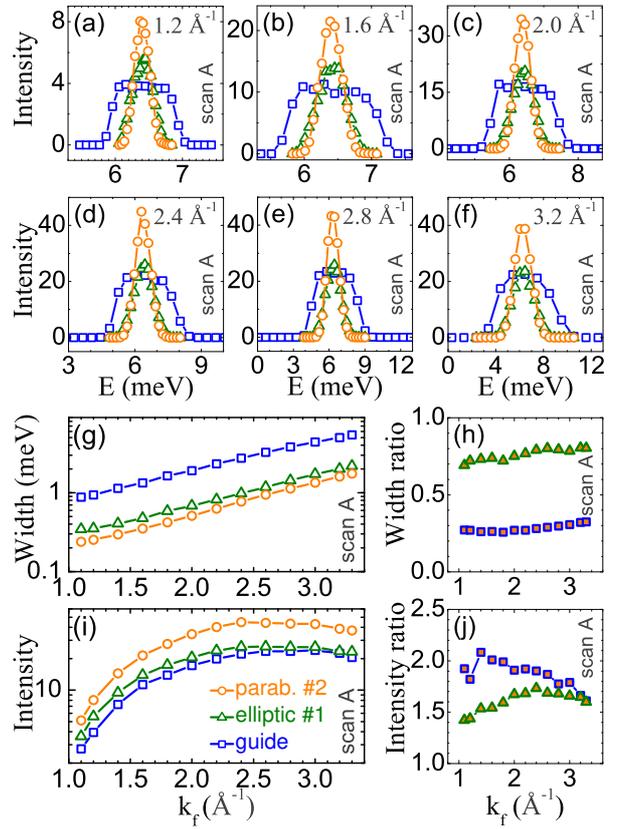}
\end{center}
\caption{(a-f) Results of constant-Q-scans of type 'scan A' at
(1~1~0) for \parB\ (\emph{orange circles}), \ellA\
(\emph{green triangles}) and \strai\ (\emph{blue squares}) for
different $k_f$. (g-j) The phonon peak widths and intensities as a
function of $k_f$ together with the ratios of the different
concepts; \emph{green/orange triangles}: ratios of \parB\ versus
\ellA\ concept, \emph{blue/orange squares}: ratios of \parB\
versus \strai\ concept.} \label{figI}
\end{figure}
\begin{figure}[!b]
\begin{center}
\includegraphics*[width=1\columnwidth,clip]{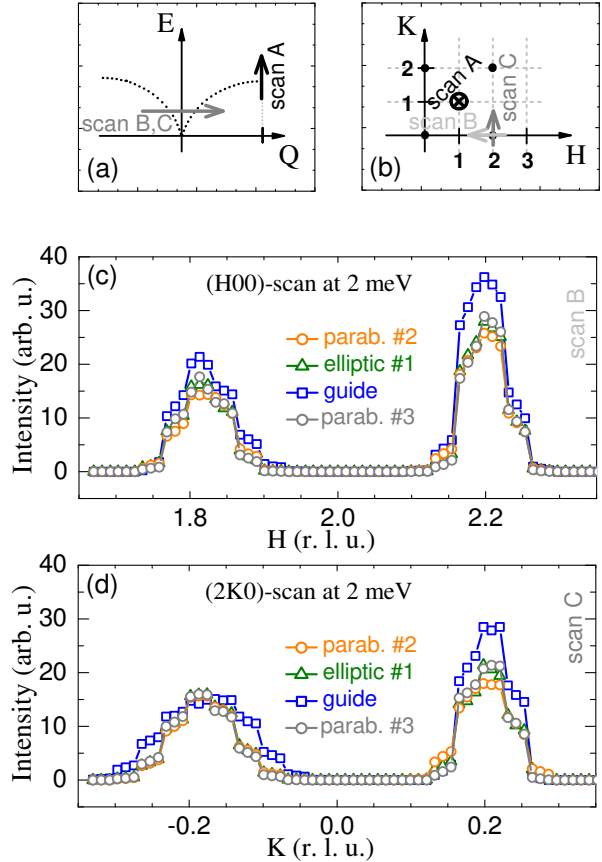}
\end{center}
\caption{(a,b) Scan directions for the three simulated phonon
scans in reciprocal space. (c) Longitudinal scan through (2~0~0)
at 2~meV with $k_f$~=~1.5~\AA$^{-1}$. (d) Transverse scan through
(200) at 2~meV with $k_f$~=~1.5~\AA$^{-1}$. \emph{Orange circles}:
\parB\ concept, \emph{green triangles}: \ellA\ concept, \emph{blue squares}:
standard \strai\ guide.} \label{figJ}
\end{figure}
Finally, we have simulated inelastic scans across transverse
and longitudinal acoustic branches of the phonon dispersion of a
\emph{fcc} single crystal using the
\emph{Phonon$_{-}$simple}-component provided by \mcs\ which is
based on the expressions of cross sections provided by
\emph{Squires} \cite{squires}. The model sample has a cylindrical
shape with 2~cm diameter and 2~cm height and the crystal lattice
has 5~\AA\ lattice parameters. A low sound velocity corresponding
to a slope of the phonon dispersion which amounts to
8~meV/\AA$^{-1}$ (at 300~K) has been used. The sample mosaic is not
included in the \emph{Phonon$_{-}$simple}-component, i.e.
$\eta$~=~0. For these simulations a double-focusing analyzer
geometry has been applied.

\par The simulated intensities of constant-Q-scans
 are shown in Fig.~\ref{figI}. The intensities have been
calculated at the detector position. In order to avoid any
focusing effects of the resolution ellipsoid with the phonon
dispersion (focusing side and defocusing side), we have studied
first the phonon dispersion at the zone boundary because its
dispersion is flat, i.e. we have simulated constant-Q-scans at
(1~1~0) in reciprocal space with varying energies around
$\sim$6.4~meV; compare also Fig.~\ref{figJ}~(a). We will denote
this type of scan by 'scan A' and for the sake of simplicity we
will only show the results for the three major concepts with
either (nearly) full or no focusing properties. In
Fig.~\ref{figI}~(a-f) the resulting energy-scans are shown for
different values of the final wave vector $k_f$. For a more
quantitative analysis the peak intensities measured at
(\textbf{Q},$\hbar\omega$) and the corresponding phonon peak
widths are shown in Fig.~\ref{figI}~(g,i). Finally, in
Fig.~\ref{figI}~(h,j) the ratios of these phonon widths and
intensities for the parabolic versus the two other models are
shown.
Obviously, the parabolic concept has the highest energy resolution
together with the highest intensities among the different focusing
concepts. The energy
resolution is 30\% better than for the elliptic concept with at least 50\% more intensity. Compared to the
standard straight concept the energy resolution is more than 70\%
better, having even 100\% more intensity in the phonon peak.
\par Comparing the elliptic and straight concepts, the elliptic alternative
has larger intensity for medium and lower values of $k_f$ but
loses this advantage of intensity versus the straight
guide at higher $k_f$. However, the
elliptic concept can still keep its advantage in the energy
resolution compared to the straight guide even at higher values of
$k_f$.
\par In order to study also the focusing effects of the resolution ellipsoid
with the phonon dispersion, we have simulated constant-energy scans
at low energies across the zone center for the different basic
concepts. The scan directions of two constant-E-scans through
(2~0~0) are indicated in Fig.~\ref{figJ}~(a,b), one of them having
a longitudinal scan-direction ('scan B')
 and the other having a transverse scan-direction ('scan C'), thus being sensitive to phonons
 with longitudinal and transverse polarization, respectively.
Both constant-E-scans have been calculated with 2~meV energy
transfer and in the constant-$k_f$ operation mode with
$k_f$~=~1.5~\AA$^{-1}$. As can be seen in Fig.~\ref{figJ}~(c,d)
there is no significant difference between elliptic (\ellA) and
parabolic (\parB) concepts
but interestingly the standard straight
guide is clearly superior to the focusing guides,
either elliptic or parabolic.
This can be attributed to the higher Q-resolution of the \strai\ guide.
\begin{figure*}[!t]
\begin{center}
\rotatebox{270}{\includegraphics*[width=1.105\columnwidth,clip]{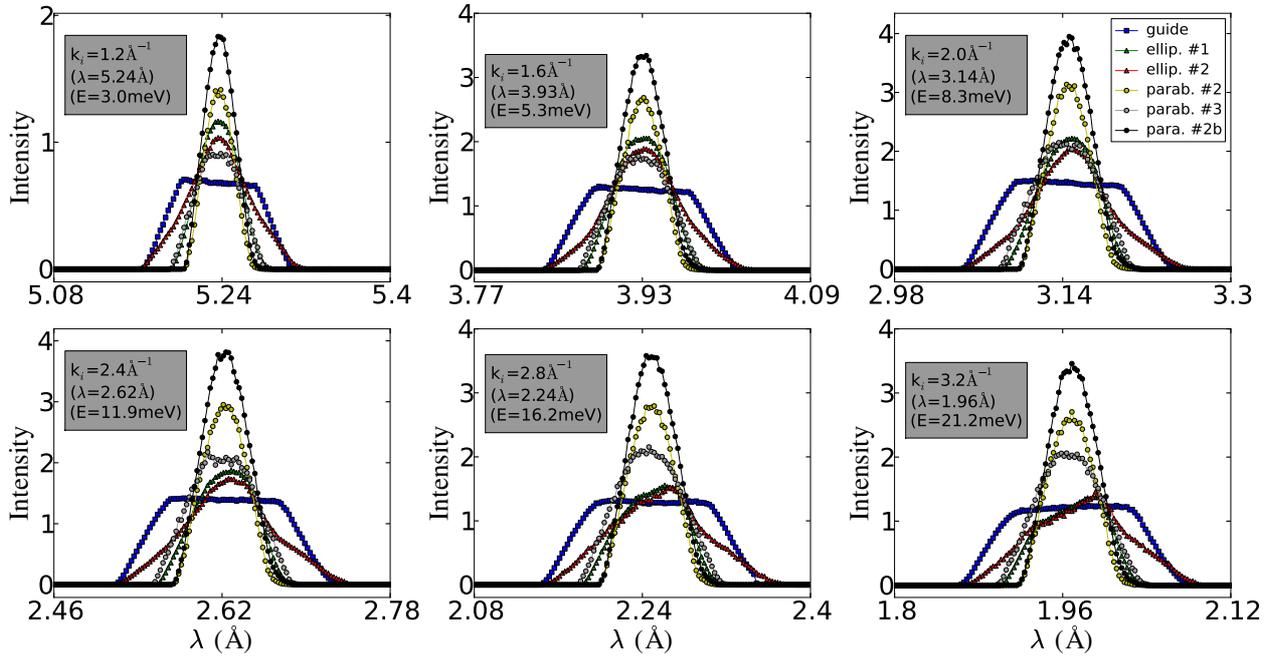}}
\end{center}
\caption{Calculated neutron scattering intensity at the sample
position for different concepts: \parB\ with optimized
monochromator, i.e. \parM\ (\emph{black circles}) versus \parB\
(\emph{yellow circles}), \parC\ (\emph{grey circles}), \ellA\
(\emph{green triangles}), \ellB\ (\emph{red triangles}) and
\strai\ (\emph{blue squares}) with the standard monochromator
setup.} \label{figA}
\end{figure*}

Complementary constant-Q-scans at
(2.2~0~0) which are not shown here, corroborate these findings -
the intensity of the \strai\ concept is clearly superior to the
two focusing concepts. However, the energy resolution (FWHM) of
the different concepts remains almost identical and amounts to
0.60(6)~meV, 0.59(7)~meV,and 0.61(6)~meV for the
\parB, \ellA and \strai\ concepts. Thus, in constant-Q scans measured
at a steep dispersion, the \parB\ concept exhibits less intensity
than the standard straight guide but not a worsened resolution.
Simulations of scans of type 'scan-B' and
'scan-C' within the
\parC\ concept show no significant improvement, see Fig.~\ref{figJ}~(c,d). However, a
larger sample mosaic will be beneficial for the
parabolic setup versus the other concepts since larger parts of
the divergence-band will interact with the sample (remember that
zero sample mosaic was used in these phonon calculations).

\section{Optimization  of the Monochromator}
Because the distance $d_{MS}$ between sample and monochromator is
rather small for the intended compact design and for focusing from
point to point (with $\Delta d$~$\sim$~0.3~m), a reduction of the
HOPG monochromator crystal size seems useful in order to approach
a truly curved surface since the curvatures become rather large.
Furthermore, an increase of the mosaic spread could be beneficial
for these purposes and also for transporting a larger part of the
divergence-bandwidth provided by the cold guide. However, the gap
between the monochromator crystals as well as the required higher
precision of the monochromator mechanics limit the reduction of the crystal size.
Enhancing the monochromator
crystal mosaic might result in a coarser resolution in
measurements with a flat monochromator setting without any
collimation. Hence, only a small crystal size reduction and a
small increase of the crystal mosaic seems to be useful. After simulating several different
setups with different crystal sizes (ranging from
10~mm~$\times$~10~mm to 20~mm~$\times$~20~mm) and with different
crystal mosaic (ranging from 0.4\grad\ to 1.0\grad), we decided to
choose a monochromator array with 14~mm~$\times$~14~mm crystal
size, i.e. a total reduction of crystal size (area) by $\sim$50\%.
Even smaller crystal size would yield higher simulated intensities
at the sample position but might result in mechanical problems of
the monochromator construction. Furthermore, a mosaic spread
enhanced by 70\% was chosen as the optimum for the intensity at
the sample position. We denote this configuration of optimized
monochromator in combination with the
\parB-concept as the '\parM'-concept. The combination of
\parC\ with this optimized monochromator is
denoted as \parMC. In Fig.~\ref{figA} the
results of the simulations for the high-E-resolution configuration
\parM\ (\emph{solid black circles}) are shown and compared with
other configurations all having the standard monochromator design
according to Ref.~\cite{ellipt}. For all calculated neutron
energies the gain of total neutron-flux at the sample position is
of the order of one third higher than the high flux of the
'conventional'
\parB\ configuration. Furthermore, the multiple peak structure transverse
to \textbf{Q} vanishes since the intensity is smeared out due
to the larger mosaic spread (see Fig.~\ref{figH}).
\begin{figure}[!h]
\begin{center}
\includegraphics*[width=1\columnwidth,clip]{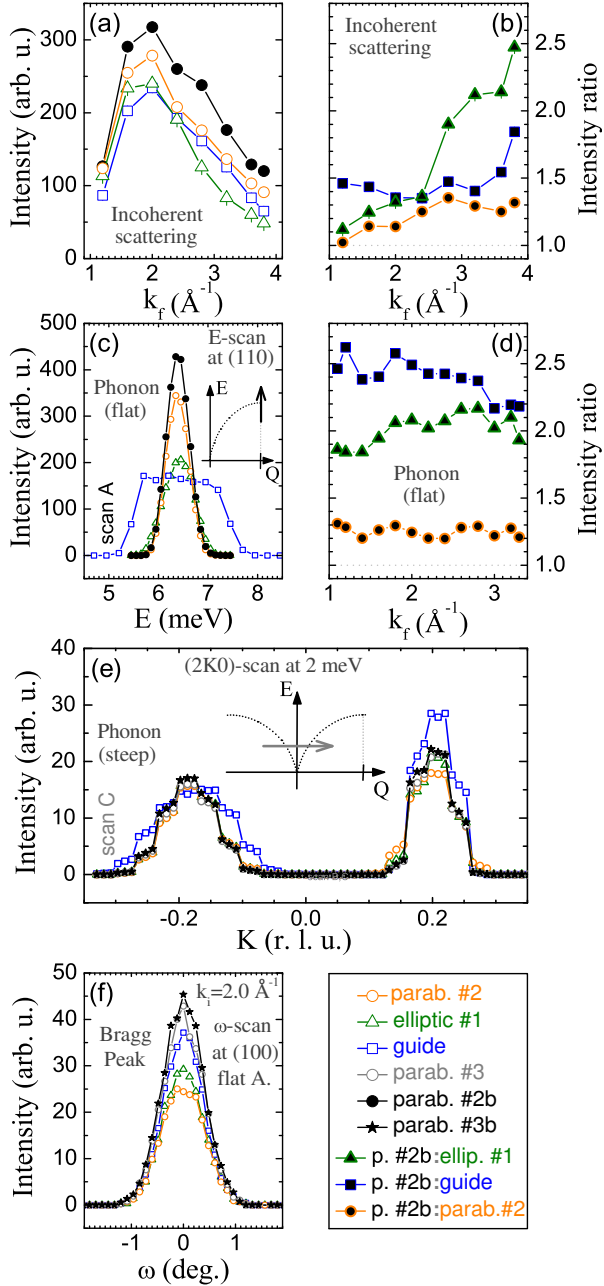}
\end{center}
\caption{(a) Intensity (amplitude) of simulated energy-scans
across the elastic line of an incoherent scatterer. In (b) the
corresponding ratios are shown; compare legend. (c) Simulated
constant-Q-scan ('scan A'). (d) Ratio of the intensity obtained
in the simulation of 'scan A' in configuration \parM\ versus \ellA, \strai\ and \parB. (e) Simulated
constant-E-scan at 2~meV ('scan C') for different models. (f) A
simulated $\omega$-scan at (1~0~0).} \label{figM}
\end{figure}
\par We have also
simulated energy scans across the elastic line of an incoherent
scatterer under the same conditions as described in the preceding
section. The resulting intensities are shown in
Fig.~\ref{figM}~(a) and the ratios of the high-E-resolution \parM\
and the other different high-E-resolution focusing concepts as
well as the standard straight guide are shown in
Fig.~\ref{figM}~(b). Thus, by choosing an optimized monochromator
design (\emph{black solid points}), a distinct increase of
intensity can be expected in inelastic measurements in general.
Fig.~\ref{figM}~(c,d) show the high performance of
\parM\ (\emph{black solid points}) for measuring flat
dispersions. However,  Fig.~\ref{figM}~(e) shows that the standard
straight guide is still superior for measuring steep dispersions.

\par Furthermore, we were also curious to see whether
any performance collapse appears in the \parM\ configuration in
elastic scans since the crystal mosaic has been enhanced.
Therefore, $\omega$-scans have been simulated under the same
conditions as described in the preceding section. In spite of
using a flat analyzer geometry there is no drop but even a small
increase of intensity for \parMC\ (\emph{solid black
stars}) as can be seen in Fig.~\ref{figM}~(f).

\par Hence, the monochromator can be
optimized for the use in double focusing geometry by choosing
crystal size and crystal mosaic properly depending on the
occurring monochromator curvatures which are larger for a fully
focusing guide than for a non-focusing straight guide.  Since
elastic scans usually do not suffer from a lack of intensity, any
problems due to a larger crystal mosaic can be handled by
using an appropriate collimation.
\par Finally, we note that a reduction of the $\Delta
d$-shift towards zero does not  yield any improvement but even a
clearly diminished intensity for 'scan C' in Fig.~\ref{figM}~(e).

\section{Conclusion}
We have performed Monte-Carlo simulations in order to
determine the optimal focusing design for a cold-neutron
triple-axis spectrometer which will be installed at the end of a
curved neutron guide (NL1 at the FRM-II). The elliptic guide \cite{ellipt}
turns out not to be the ideal choice since the elliptic
configuration focuses also neutrons with higher divergence into
its focal point. In contrast, the parabolic alternative yields much higher intensities
by focusing neutrons with small divergence into its focal point.
\par The highly focusing design of the parabolic
concept also results in a distinctly better energy resolution and
does not suffer that much from a severe
multi-peak structure transverse to \textbf{Q} like the elliptic
concept. Hence, intensity as well as energy and \textbf{Q}-resolution are superior
in the parabolic concept. This improved performance of parabolic focusing could be also
verified in simulations for phonon-scans and of scans across the
elastic line of an incoherent scatterer.
\par However, for steep dispersions and low neutron energies,
both focusing concepts either elliptic or parabolic become quite
similar having a significant intensity disadvantage compared to a
standard straight guide.
\par The monochromator is the essential element of the whole
focusing concept and can be optimized for the required curvatures
in the double-focusing operation mode. If all distances are small
and the curvatures become large, a smaller crystal size and a
slightly enhanced crystal mosaic are beneficial in order to follow
the large curvatures and probably also to transport a larger part
of the divergence bandwidth provided by the guide.
\par We may conclude that for an instrument at the end of a neutron guide,
the parabolic concept is clearly superior to the elliptic concept.
In most cases the parabolic concept is distinctly better than a
standard straight guide unless the measured dispersion is steep
and high Q-resolution is desired.

\section{Acknowledgements}
This work was supported by the German Federal Ministry of
Education and Research (BMBF) by project 05KN7PK1. We thank
A.~Ostermann for providing the model of the cold source and the
cold guide NL-1 of the FRM-II up to the upstream instrument \nrex.
We thank M.~Janoschek for providing the optimized guide parameters
of the elliptic guide.

\bibliographystyle{elsarticle-num-names}

\bibliography{KompassFoc}
\end{document}